\documentclass[aps,prb,reprint,tightenlines,superscriptaddress]{revtex4-2} 

\usepackage{amsmath,amsfonts,amssymb}
\usepackage{mathrsfs}
\usepackage{xparse}
\usepackage{physics}
\usepackage{booktabs}
\usepackage{graphicx}
\usepackage[utf8]{inputenc}
\usepackage[english]{babel}
\usepackage[colorlinks = true,
            linkcolor = blue,
            urlcolor  = blue,
            citecolor = blue,
            anchorcolor = blue]{hyperref}
\usepackage{subfigure}
\usepackage{textcomp}
\DeclareUnicodeCharacter{0301}
\renewcommand{\mathbf}{\boldsymbol}
\newcommand*\DS{\textcolor{blue}}
\DeclareDocumentCommand\term{mmg}{%
  {$^{#1}$#2%
  \IfNoValueF {#3} {$_{#3}$}%
  }%
}

\begin{document}

\newcommand{\Er}{Er\textsuperscript{3+}}
\newcommand{\Nd}{Nd\textsuperscript{3+}}
\newcommand{\YO}{Y\textsubscript{2}O\textsubscript{3}}
\newcommand{\ErYO}{Er\textsuperscript{3+}:Y\textsubscript{2}O\textsubscript{3}}
\newcommand{\NdYO}{Pr\textsuperscript{3+}-Nd\textsuperscript{3+}:Y\textsubscript{2}O\textsubscript{3}}
\newcommand{\Ertransition}{${}^4I_{13/2}-{}^4I_{15/2}{\,}$}
\newcommand{\Ertransitionzero}{${}^4I_{13/2}(0)-{}^4I_{15/2}(0){\,}$}
\newcommand{\Ndtransition}{${}^4F_{3/2}-{}^4I_{9/2}{\,}$}
\newcommand{\Ndtransitionzero}{${}^4F_{3/2}(0)-{}^4I_{9/2}(0){\,}$}

\title{Optical coherence properties of Kramers' rare-earth ions at the nanoscale for quantum applications}

\author{Mohammed K.\ Alqedra}
\thanks{These authors contributed equally.}
\affiliation{Department of Physics, Lund University, P.O. Box 118, SE-22100 Lund, Sweden}
\author{Chetan Deshmukh}
\thanks{These authors contributed equally.}
\affiliation{ICFO-Institut de Ciencies Fotoniques, The Barcelona Institute of Science and Technology, 08860 Castelldefels, Barcelona, Spain}

\author{Shuping Liu}
\affiliation{Chimie ParisTech, PSL University, CNRS, Institut de Recherche de Chimie Paris, 75005 Paris, France}
\affiliation{Shenzhen Institute for Quantum Science and Engineering, Southern University of Science and Technology, 518055 Shenzhen, China}

\author{Diana Serrano}
\affiliation{Chimie ParisTech, PSL University, CNRS, Institut de Recherche de Chimie Paris, 75005 Paris, France}

\author{Sebastian P.\ Horvath}
\affiliation{Department of Physics, Lund University, P.O. Box 118, SE-22100 Lund, Sweden}
\author{Safi Rafie-Zinedine}
\affiliation{Department of Physics, Lund University, P.O. Box 118, SE-22100 Lund, Sweden}
\author{Abdullah Abdelatief}
\affiliation{Department of Physics, Lund University, P.O. Box 118, SE-22100 Lund, Sweden}
\author{Lars Rippe}
\affiliation{Department of Physics, Lund University, P.O. Box 118, SE-22100 Lund, Sweden}
\author{Stefan Kr\"oll}
\affiliation{Department of Physics, Lund University, P.O. Box 118, SE-22100 Lund, Sweden}

\author{Bernardo Casabone}
\affiliation{ICFO-Institut de Ciencies Fotoniques, The Barcelona Institute of Science and Technology, 08860 Castelldefels, Barcelona, Spain}

\author{Alban Ferrier}
\affiliation{Chimie ParisTech, PSL University, CNRS, Institut de Recherche de Chimie Paris, 75005 Paris, France}
\affiliation{Facult{\'e} des Sciences et Ing{\'e}nierie,  Sorbonne Universit{\'e}, UFR 933, 75005 Paris, France}

\author{Alexandre Tallaire}
\affiliation{Chimie ParisTech, PSL University, CNRS, Institut de Recherche de Chimie Paris, 75005 Paris, France}

\author{Philippe Goldner}
\affiliation{Chimie ParisTech, PSL University, CNRS, Institut de Recherche de Chimie Paris, 75005 Paris, France}

\author{Hugues de Riedmatten}
\affiliation{ICFO-Institut de Ciencies Fotoniques, The Barcelona Institute of Science and Technology, 08860 Castelldefels, Barcelona, Spain}
\affiliation{ICREA-Instituci{\'o} Catalana de Recerca i Estudis Ava\c{c}ats, 08015 Barcelona, Spain}

\author{Andreas Walther}
\affiliation{Department of Physics, Lund University, P.O. Box 118, SE-22100 Lund, Sweden}

\date{\today}

\begin{abstract}

Rare-earth (RE) ion doped nano-materials are promising candidates for a range of quantum technology applications. Among RE ions, the so-called Kramers' ions possess spin transitions in the GHz range at low magnetic fields, which allows for high-bandwidth multimode quantum storage, fast qubit operations as well as interfacing with superconducting circuits. They also present relevant optical transitions in the infrared. In particular, \Er{} has an optical transition in the telecom band, while \Nd{} presents a high-emission-rate transition close to 890 nm. In this paper, we measure spectroscopic properties that are of relevance to using these materials in quantum technology applications. We find the inhomogeneous linewidth to be 10.7~GHz for \Er{} and 8.2~GHz for \Nd{}, and the excited state lifetime $T_1$ to be 13.68~ms for \Er{} and 540~$\mu$s for \Nd{}. We study the dependence of homogeneous linewidth on temperature for both samples, with the narrowest linewidth being 379~kHz ($T_2 = 839$~ns) for \Er{} measured at 3~K, and 62~kHz ($T_2 = 5.14$~$\mu$s) for \Nd{} measured at 1.6~K. Further, we investigate time-dependent homogeneous linewidth broadening due to spectral diffusion and the dependence of homogeneous linewidth on magnetic field, in order to get additional clarity of mechanisms that can influence the coherence time. In light of our results, we discuss two applications: single qubit-state readout and a Fourier-limited single photon source.
 
\end{abstract}

\maketitle
\section{Introduction}
Rare Earth (RE) ions doped into inorganic crystals have been widely investigated as a solid-state platform for quantum applications, including as quantum memories \cite{hedges_2010,Zhong2017Sep,Seri2017,Ortu2022Mar}, quantum information processors \cite{Longdell2004,GOLDNER20151,Ahlefeldt_2020,Kinos2022}, quantum sensors \cite{Thorpe2011,Degen2017} and 
microwave-to-optical transducers \cite{Bartholomew2020, Williamson2014}. This interest is mainly motivated by the exceptionally long optical and spin coherence times of the 4f-4f transitions at cryogenic temperatures. Other motivations include the highly dense spectral storage capabilities, which are facilitated by narrow homogeneous lines (kHz) spread over much wider inhomogeneous lines (GHz). This allows the different frequency channels to be used to address the spin states of different ions, even though they can be within nanometers of each other, giving a high potential qubit density \cite{Equall1994,Thiel2011,Zhong2015Jan}. High-fidelity detection of single RE ions is important for several applications, such as quantum networks and quantum computing \cite{Asadi2018Sep,DiVincenzo2000Sep}. However, the low spontaneous emission rate of RE ions resulting from the dipole-forbidden nature of the 4f-4f transitions makes this very challenging.

During the last few years, single RE ion detection has been demonstrated by Purcell enhancing the weak 4f transitions using different types of cavities \cite{Dibos2018, Chen2020Oct, Zhong2018, Kindem2020Apr, Casabone2018Sep}. To incorporate RE ion doped crystals into small mode-volume optical cavities, nano-structuring of the host crystal is essential. In this regard, yttrium oxide (\YO{}) has emerged as a very promising host material due to the excellent coherence properties that RE ions doped into nanoscale \YO{} have demonstrated \cite{Harada2020Aug, Liu2020Aug, Fukumori2020, Serrano2019, Bartholomew2017Feb}. Recently, RE ions doped into \YO{} nanoparticles were also coupled to high-finesse fibre based micro-cavities to demonstrate Purcell-enhanced emission \cite{Casabone2018Sep, Casabone2021}.

Among RE ions, so-called Kramers' ions with an odd number of 4f electrons, including erbium (\Er{}), neodymium (\Nd{}) or ytterbium (Yb$^{3+}$), offer unique possibilities due to their high magnetic-moments. Their spin transitions can be tuned in the GHz range even at low magnetic fields, which can be used for high-bandwidth multi-mode quantum storage \cite{Jobez2016Mar,Laplane2017May}, fast qubit operations \cite{Chen2020Oct,Ruskuc2021Aug}, and for interfacing with superconducting qubits \cite{Fernandez2015Dec,Dold2019May}. 
 
In this paper, we investigate two of the most appealing Kramers' ions for quantum applications, \Er{} and \Nd{}. The \Ertransition{} transition in \Er{} at 1.5 $\mu$m lies in the telecommunication band, and hence offers easy integration into existing commercial fiber-optic networks for the purpose of quantum communication. The \Ndtransition{} transition in \Nd{} has one of the highest oscillator strengths among all RE ions and a high branching ratio, which together make it very promising to achieve high cooperativity in an optical cavity \cite{Sun2002Jul,Cui2011Apr}.

In this work, we experimentally characterize the optical spectroscopic properties of the \Ertransition{} transition in \Er{} and  of the \Ndtransition{}  transition in \Nd{} doped into \YO{} nanoparticles. In Sec.~\ref{sec:lifetime}, we report excited state lifetimes ($T_1$) as well as inhomogeneous linewidths. In Sec.~\ref{sec:temp}, we characterize \Er{} and \Nd{} homogeneous linewidths as a function of temperature. In Sec.~\ref{sec:3ppe}, we measure time-dependent spectral diffusion in \Er{}. In Sec.~\ref{sec:bfield}, we investigate the dependence of homogeneous linewidth of \Nd{} on magnetic field. Finally, in light of our results, we present an outline for two possible quantum applications in Sec.~\ref{sec:applications}, namely Fourier-limited single photon sources and single qubit state readout. 

\section{Experiment}\label{sec:experiment}

\subsection{Material}\label{sec:material}

\begin{figure}[ht]
\centering
\includegraphics[width=1.0\columnwidth]{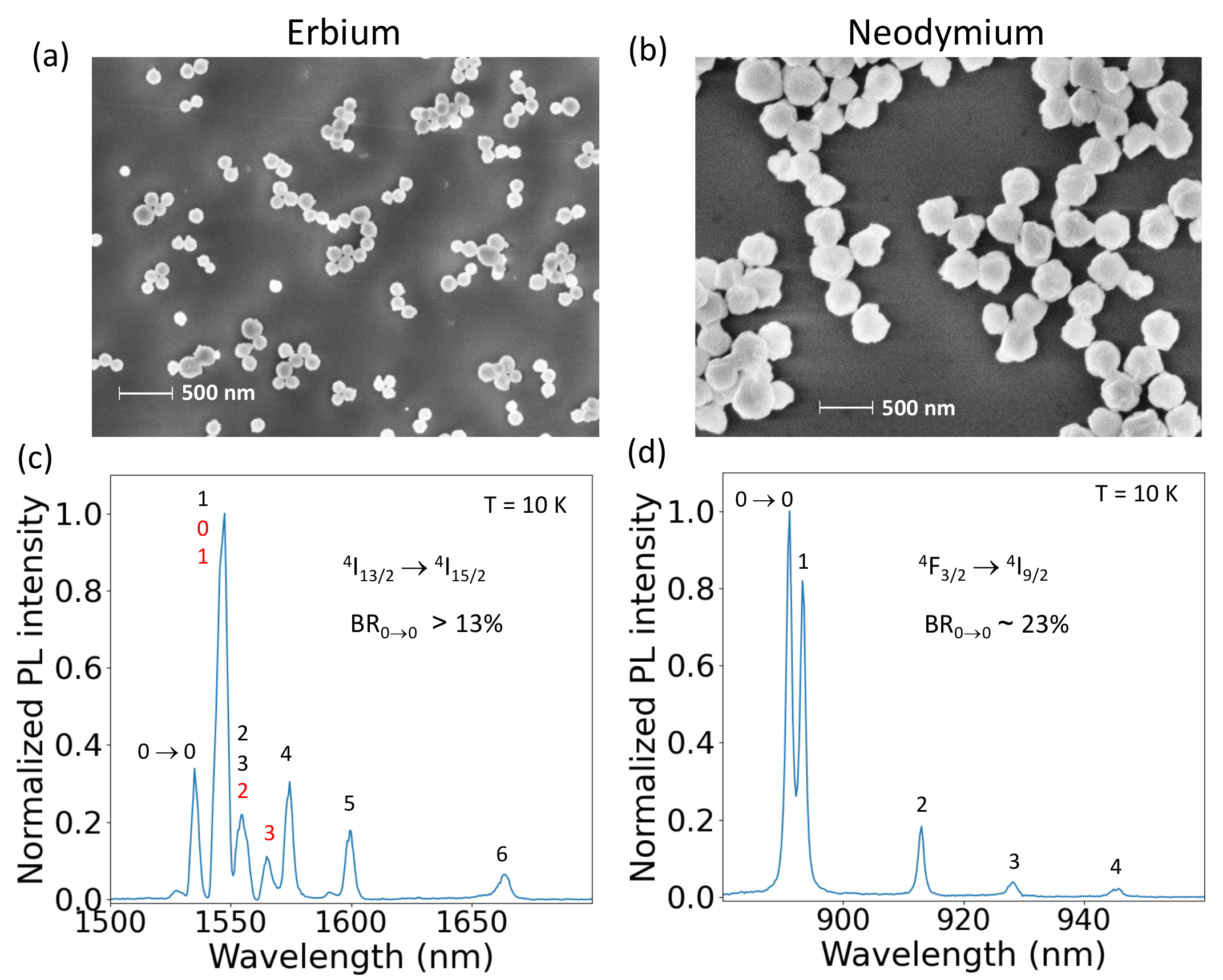}
\caption{(a) Scanning electron microscopy (SEM) image of \ErYO{} nanoparticles with average size of 150 $\pm$ 13 ~nm, and of (b) \NdYO{} nanoparticles with average size of 380 $\pm$ 24~nm. (c) Fluorescence spectra of \ErYO{} at 10~K on the \Ertransition{} transition obtained by exciting the $^4I_{11/2}$ level at 975 nm. The different peaks in the spectra were assigned to transitions from the lowest energy level in the $^4I_{13/2}$ multiplet to the different energy levels in the ground-state multiplet ($^4I_{15/2}$) for both site 1 (black) and site 2 (red). Peaks at 1548 nm and 1555 nm contain several unresolved emission lines. We estimate a lower bound branching ratio of 13~\% for the 0 $\to$ 0 line, at 1535~nm. (d) \NdYO{} on the \Ndtransition{} transition obtained by exciting the $^4F_{5/2}$ level at 813 nm, where we measure a branching ratio of 23~\% for the 0 $\to$ 0 line at 892~nm.}
\label{fig:nano}
\end{figure}

Two batches of \YO{} nanoparticles were investigated in this study: one doped with \Er{} (200 ppm at.), presenting an average particle size of 150 nm, and a second one co-doped with \Nd{} (100 ppm at.) and Pr$^{3+}$ (500 pm at.) ions, presenting average particle size of 380 nm (Fig. \ref{fig:nano} (a)-(b)). The nanoparticles were obtained by homogeneous precipitation followed by high-temperature annealing. The synthesis route is described in full details in \cite{Liu2020,Goldner2015}. The annealing temperature was set to 900 $^\circ$C for 6 h for the \Er{} sample while it was 1200 $^\circ$C for 6 h in the \Nd{}-Pr$^{3+}$ one. Lower temperature was used for \Er{} to avoid aggregation during annealing due to the nanoparticles' reduced size. In addition, \Er{} nanoparticles were processed under pure $O_2$ plasma twice for three minutes to cure defects \cite{Liu2020}.  Particle size, morphology, and dispersion \DS{were} assessed by scanning electron microscopy (SEM) (Fig. \ref{fig:nano} (a)-(b)). The microstructure of the nanoparticles was determined by X-ray diffraction (XRD) analysis using the Williamson-Hall method, as poly-crystalline, with crystallites of 58 $\pm$ 11 nm for \Er{} and 102 $\pm$ 10 nm for \Nd{}-Pr$^{3+}$ \YO. The latter also confirms that the nanoparticles present pure cubic phase with $Ia$-3 space group (JCPDS 01-080-6433). In this structure, RE ions occupy two different crystal sites, with $C_2$ and $S_6$ point symmetries respectively. In this work, we will investigate spectroscopic properties of \Er{} and \Nd{} ions in $C_2$ sites. 

\YO{} ceramics doped with \Er{} (200 ppm at.) and \Nd{} (100 ppm at.) were also synthesized by solid state reaction to be used as bulk references. Those were fabricated from a mix of commercial \YO{} and Er$_2$O$_3$ (or Nd$_2$O$_3$) oxide powders (with 99.99~\% purity). The powders were pressed into pellets with a pressure of 5 MPa, and then annealed at 1500~$^{\circ}$C for 48 h in an air atmosphere. The obtained ceramics were cut into thin slabs of $\sim$ 260 $\mu$m thick for the optical measurements.

\subsection{Procedures}\label{sec:procedures}

For measurements on \Er{}, the transition of interest is \Ertransitionzero{} at around $1535$~nm, which lies in the telecom band. The excitation light was obtained from an external cavity diode laser at $1535$~nm, which was then amplified by an Erbium-doped Fibre Amplifier to reach a peak power of $\sim500$~mW. The amplified light was sent through a double-pass Acousto-Optic Modulator (AOM) for pulse modulation. The light was then filtered with a bandpass filter at $1535 \pm 3$~nm before being focused onto the sample using a $35$~mm spherical lens. Typical peak power at the sample was $\sim150$~mW. The samples were prepared on a copper holder with a pinhole and placed on the cold-finger of a closed-cycle cryostat, which was cooled down to a minimum of $3$~K. The temperature of the sample could be increased with the help of heaters installed on the cold finger. The scattered light after the sample was then loosely focused onto an Avalanche Photo-Diode (APD) with the help of two spherical lenses, each with a focal length of $60$~mm. The signal from the APD was recorded on an oscilloscope. For two-pulse photon echo measurements, the two pulses had a duration of 250~ns in length, and the delay between them was varied from 150~ns to 3~$\mu$s. For three-pulse photon echo measurements, all the three pulses were 250~ns, and the delay between the first two pulses was kept constant at 150~ns while the delay between the second and third pulse was varied from 3~$\mu$s to 40~$\mu$s. A heterodyne scheme was used to detect the echo where the heterodyne pulse was $1$~$\mu$s long with a frequency detuning of $24$~MHz. The resulting beat note signal from the APD was recorded on an oscilloscope after passing through a high-pass filter with a cut-off frequency of $\sim 5$~MHz to filter out electronic noise. A small magnetic field of $\sim 100$~G was applied for all of the measurements with the help of permanent magnets.

For measurements on \Nd{}, a CW Ti:sapphire laser operating at a wavelength of $892$~nm was used to target the \Ndtransitionzero{} transition. A double pass AOM was also utilized for pulse shaping. In order to calibrate for laser intensity fluctuations, 10\% of the input was focused into a reference detector before the sample. The rest of the light, which amounted to $\sim10$ mW, was focused on the sample. A copper mount was used to hold the sample inside a bath cryostat operating at 2.15 K. The cryostat was equipped with a superconducting magnet system that can reach up to 7 T. For the temperature dependence measurement, the cryostat was configured to cool using Helium exchange gas. The scattered light after the sample was collected and focused onto an APD. For inhomogeneous line and lifetime measurements, a $900$~nm long-pass filter was mounted before the detector to filter out the excitation light while passing part of the fluorescence emission (see Fig.~\ref{fig:nano}(d)). For coherence measurements, a two-pulse photon echo sequence was used with 400-ns-long pulses and direct echo detection. 

Low temperature emission spectra shown in Fig.~\ref{fig:nano}(c)-(d) were recorded at 10 K using a closed-cycle cryostat. Excitation was carried out with a tunable optical parametric oscillator (OPO) pumped by a Nd$^{3+}$:YAG Q-switched laser (Ekspla NT342BSH with 6 ns pulse length and 10 Hz repetition rate). Spectra were recorded using an Acton SP2300 spectrometer equipped with the following visible and infrared gratings: 300 grooves/mm centered at 1200 nm, 300 grooves/mm centered at 500 nm, and 600 grooves/mm centered at 1.2 $\mu$m. An ICCD camera (Princeton Instruments) and a Ge detector were used to detect Nd$^{3+}$ and Er$^{3+}$ emissions respectively. For \Er{}, the excitation wavelength was set to 975~nm while emissions from the \Ertransition{} transition were collected. We note that the spectrum in Fig.~\ref{fig:nano}(c) contains lines from both $C_2$ and $S_6$ sites. We estimate a branching ratio $>$ 13~\% on the 0 $\to$ 0 line for \Er{} ions in $C_2$, with the 0 denoting the lowest crystal field level of a given J multiplet. This is however a lower bound value as not all $S_6$ emission lines could be resolved from that of $C_2$. For \Nd{}, the excitation wavelength was set to 813~nm while the collection was carried out between 880 and 940~nm on the \Ndtransition{} transition. We estimate a branching ratio of 23~\% for the 0 $\to$ 0 line from the spectrum in fig. ~\ref{fig:nano}(d), taking into account the branching ratio for the \Ndtransition{} transition given in literature \cite{Walsh:02}. In contrast to \Er{}, there is no emission from \Nd{} ions in $S_6$ sites in \YO.

\section{Results}\label{sec:results}

\subsection{Inhomogeneous linewidth and lifetime}\label{sec:lifetime}

\begin{figure}[h!]
\centering
\includegraphics[width=1.0\columnwidth]{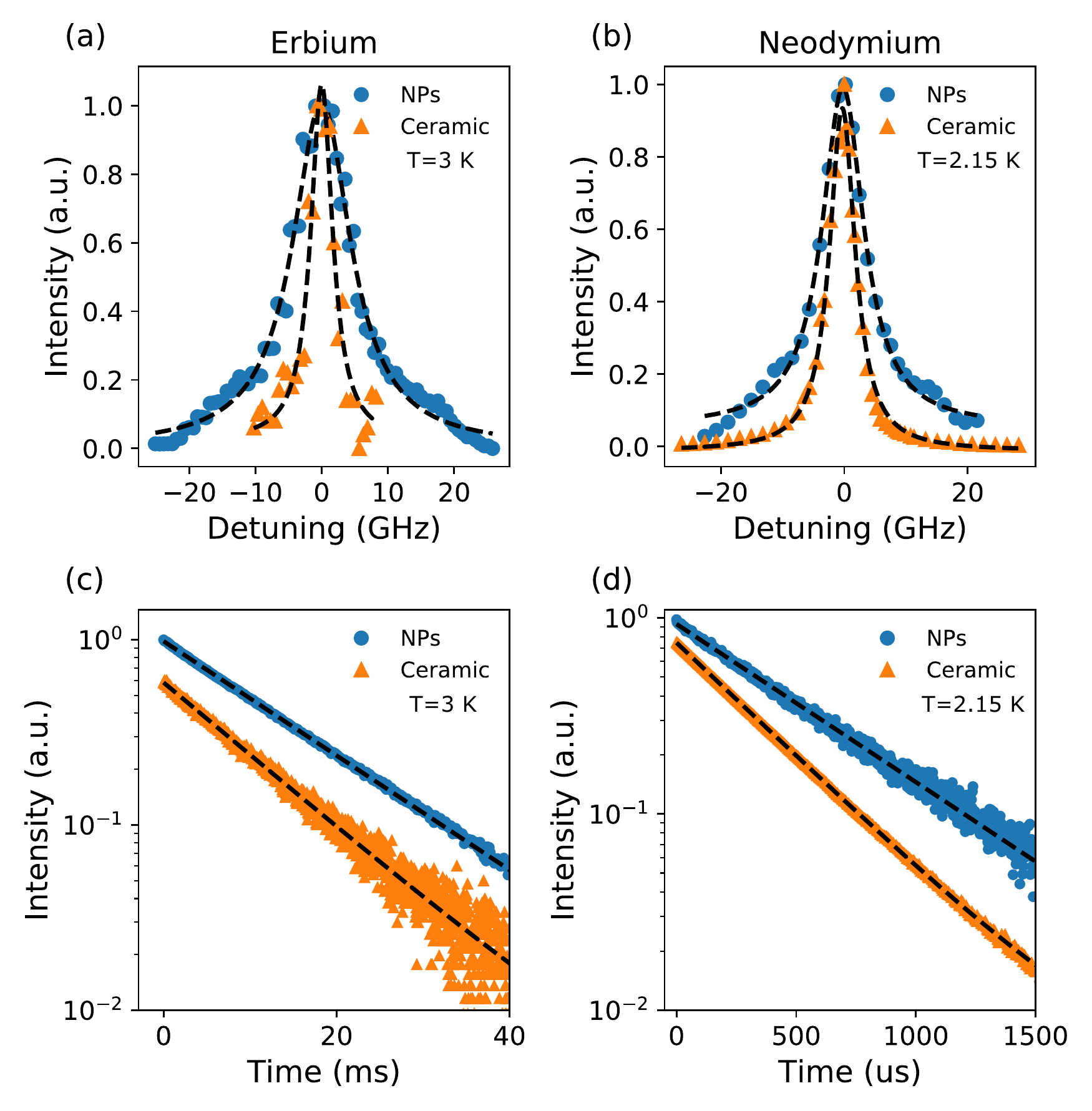}
\caption{Inhomogeneous linewidth of the 0 $\to$ 0 line of (a) \Er{} was measured to be 3.98~GHz in ceramic (orange triangles), and 10.7~GHz in nanoparticles (blue circles); and that of (b)\Nd  was measured to be 5.03~GHz in ceramic (orange triangles), and 8.20~GHz in nanoparticles (blue circles). Lifetime of the excited state of (c) \ErYO{} was measured to be 10.91~ms in ceramic (orange triangles), and 13.68~ms in nanoparticles (blue circles); and that of (c) \Nd was measured to be 373~$\mu$s in ceramic (orange triangles), and 540~$\mu$s in nanoparticles (blue circles).}
\label{fig:lifetime}
\end{figure}

The optical inhomogeneous lines for all the samples are shown in Fig.~\ref{fig:lifetime}(a)-(b). For \ErYO{}, the center of the inhomogeneous line in the ceramic sample was found to be at 1535.43~nm with a linewidth of 3.98~GHz, while in the nanoparticles the line was red-shifted to 1535.48~nm, with a linewidth of 10.7~GHz. This goes along with a tensile strain of 2.3$\times10^{-4}$ derived from the Williamson-Hall analysis for \ErYO{} nanoparticles. The observed tensile strain and spectral red-shift could be here due to the lower annealing temperature applied to the sample (900 $^\circ$C), as lower temperatures are often less efficient in curing oxygen-vacancy-type defects \cite{Liu2020}. For \Nd{}:\YO{}, the center of the inhomogeneous line in the ceramic sample was found to be at 892.17~nm with a linewidth of 5~GHz, while in the nanoparticles the line shows a blue-shift to 892.16~nm, with a linewidth of 8.2~GHz. XRD spectrum analysis shows a small compressive strain of -2.5$\times10^{-5}$ in this case. The codoping with Pr$^{3+}$ (500 ppm), present in the nanoparticles sample while not in the ceramic is in most likelihood responsible for the small blue-shift and compressive strain. The codoping can also explain the broadening of the inhomogeneous linewidth in the \NdYO{} nanoparticles \cite{Serrano2016}. For the \ErYO{} nanoparticles, the broadening is again most likely related to a larger amount of defects in the sample associated to the lower annealing temperature and additional $O_2$ plasma processing \cite{Liu2020}.

Fig.~\ref{fig:lifetime}(c)-(d) show excited state lifetime measurements for all the samples. For \ErYO{}, the excited state lifetime measured at 3 K was found to be 10.91~ms in ceramic and 13.68~ms in nanoparticles, while for \Nd{:\YO} at 2.15 K it was 373~$\mu$s in the ceramic and 540~$\mu$s in the co-doped nanoparticles. The reduction of the spontaneous emission rate in the nanoparticles compared to ceramic can be explained by the reduction of the effective refractive index surrounding the nanoparticle samples. This agrees with previous work where an increase in the spontaneous emission rate of Eu$^{3+}$ doped nanoparticles was found when it was embedded in a PMMA layer as compared to no embedding \cite{Oliveira2015,Casabone2018Sep}.

\subsection{Homogeneous linewidth versus temperature}\label{sec:temp}

\begin{figure}
\centering
\includegraphics[width=1.0\columnwidth]{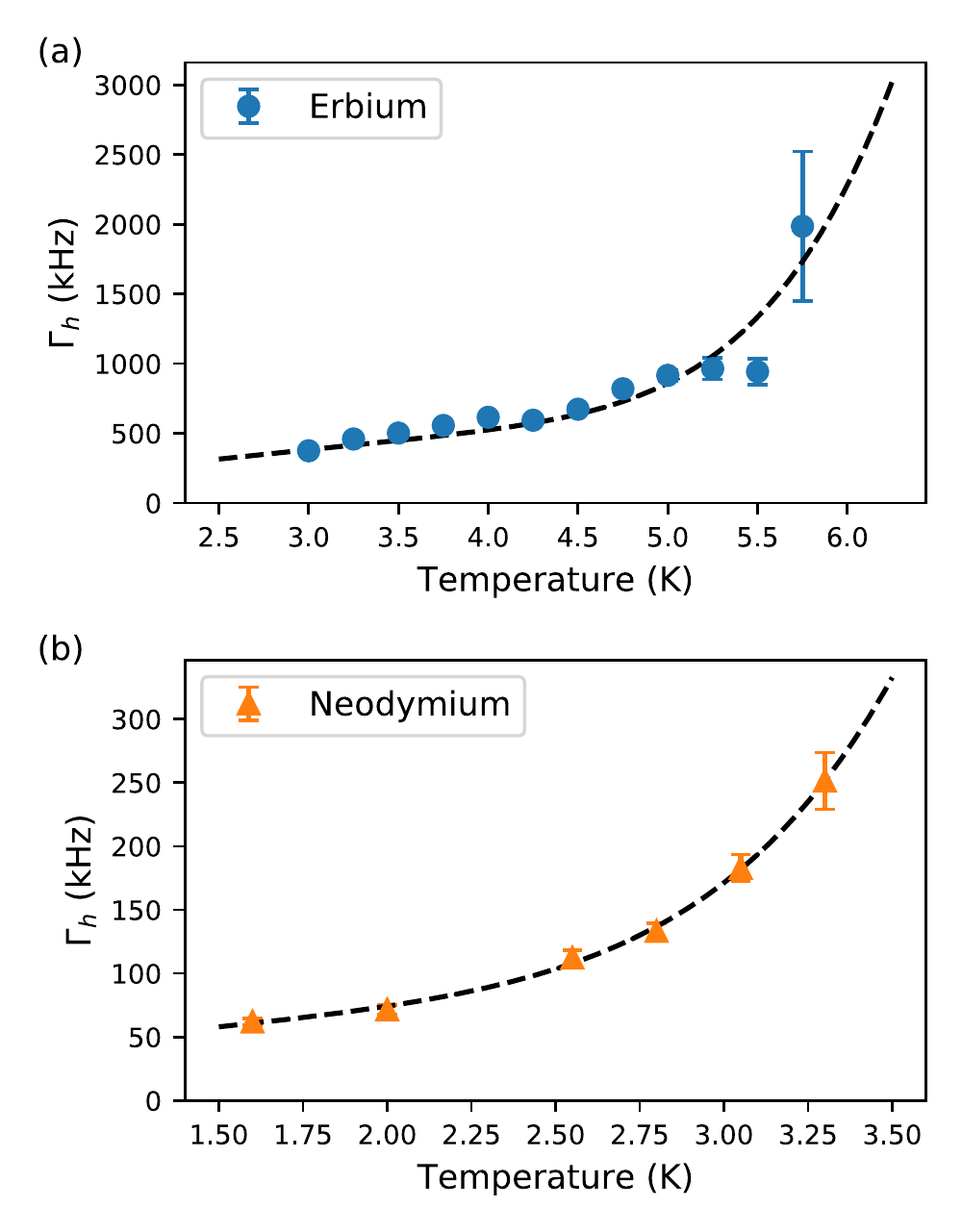}
\caption{(a) Homogeneous linewidth of \ErYO{} nanoparticles measured as a function of temperature at a magnetic field of $100$~G. Dashed line is a fit to Eq.(\ref{eq:temp}) that gives: $\Gamma_0 = 9.8$~kHz, $\alpha_{\mathrm{TLS}} = 121.7$~kHz/K, and $\alpha_{\mathrm{M}} = 11$~GHz. (b) Homogeneous linewidth of \Nd{} in \NdYO{} nanoparticles measured as a function of temperature at zero magnetic field. Fit to Eq.~(\ref{eq:temp2}) gives an upper bound of 34~kHz for $\Gamma_0$, $\alpha_{\mathrm{TLS}} = 40$~kHz/K, and $\alpha_{\mathrm{M}} = 2.7$~Hz/K$^9$. Error bars represent 95\% confidence interval obtained from the fit.}

\label{fig:temp}
\end{figure}

The homogeneous linewidth of all the samples were measured by recording the echo decay in a two-pulse photon echo sequence (see Sec.~\ref{sec:procedures} for details). For \Er{}, we measured a homogeneous linewidth of 68~kHz ($T_2=4.65~\mu$s) in the ceramic sample, while for nanoparticles it was 379~kHz ($T_2=839$~ns), both at a temperature of 3~K and a magnetic field of 100~G. This indicates that dephasing in the nanoparticles is dominated by defects and surface related noise as seen in Eu$^{3+}$ doped nanoparticles \cite{Bartholomew2017,Liu2020}. We then investigated the dependence of the homogeneous linewidth as a function of temperature, as shown in Fig.~\ref{fig:temp}(a). For $T<4.5$ K, a linear behavior is observed, evidencing interactions of Er$^{3+}$ ions with Two-Level Systems (TLS)  which are thought to be due to residual disorder in nanocrystals \cite{Bartholomew2017}.  At higher temperatures a stronger temperature dependence takes place. It can be due to changes in ground or excited state populations by 1-phonon absorption or spin-lattice relaxation (SLR), magnetic noise or elastic Raman scattering (ERS) acting directly on the optical transition \cite{Goldner2015}. In the temperature range considered, the dominant Er$^{3+}$ SLR is expected to be an Orbach process associated with crystal field splittings between lowest levels  $\Delta E$ of 39 and 31 cm$^{-1}$ for the ground and excited multiplets respectively \cite{changOpticalSpectraEnergy1982}. 
Assuming that magnetic noise is due to Er$^{3+}$ ground state spins themselves, all temperature-dependent contributions to $\Gamma_h$ thus exhibit an exponential behavior except for the ERS one which is proportional to $T^7$ at low temperatures \cite{macfarlaneHighresolutionLaserSpectroscopy2002}. 
Experimental $\Gamma_h$ values could be fitted to the expression:
\begin{equation}
\Gamma_h(T) = \Gamma_0 + \alpha_{TLS}T+ \alpha_{ERS}T^7+ \frac{\alpha_{M}}{exp(\frac{\Delta E}{k_b T})-1}
\label{eq:temp}
\end{equation}
where $k_B$ is the Boltzmann constant. With $\Delta E = 39$ cm$^{-1}$, we obtained  $\Gamma_0 = 9.8$~kHz, $\alpha_{\mathrm{TLS}} = 121.7$~kHz/K, and $\alpha_{\mathrm{M}} = 11$~GHz, with negligible contribution from $\alpha_{\mathrm{ERS}}$. As spin lifetimes of 10s of $\mu$s can be estimated from Er:Y$_2$SiO$_5$ parameters at 5 K \cite{kurkinEPRSpinlatticeRelaxation1980}, which corresponds to homogeneous broadening of 10s of kHz only, we attribute the second term of Eq. \ref{eq:temp} to magnetic noise driven by Er$^{3+}$ ground state spin relaxation. 

For \Nd{} in the co-doped particles, we measured a homogeneous linewidth of 62~kHz ($T_2=5.14 ~\mu$s) for nanoparticles at a temperature of 1.6~K with no magnetic field. The narrower linewidth compared to the \Er{} doped particles is attributed to the \Nd{} larger particle size and higher annealing temperature which decrease broadening related to defects \cite{Liu2020}. This leads to $\Gamma_h$ values for \Nd{} similar to those measured in Eu$^{3+}$ doped particles synthesized under similar conditions \cite{Bartholomew2017}.  Temperature dependence of the homogeneous linewidths in the 1.6 - 3.25 K range is presented in Fig. \ref{fig:temp}(b). Its analysis followed the one presented above for \Er{} nanoparticles, since Pr$^{3+}$ is non paramagnetic and should not have a significant effect of \Nd{} homogeneous broadening. However, as the maximum investigated temperature was only 3.25 K compared to \Nd{} ground and excited splittings of 29 and 196 cm$^{-1}$ (20 and 135 K respectively), SLR should be dominated by a two-phonon Raman process proportional to $T^9$ \cite{Scott1962} and one-phonon absorption negligible. Moreover, the ERS term at low temperatures should also be negligible \cite{Bartholomew2017}. Best fits to the data were obtained using the equation: 
\begin{equation}
\Gamma_h(T) = \Gamma_0 + \alpha_{TLS}T+ \alpha_{M}T^9
\label{eq:temp2}
\end{equation}
and gave an upper bound for $\Gamma_0$ of 34~kHz, $\alpha_{\mathrm{TLS}} = 40$~kHz/K, and $\alpha_{\mathrm{M}} = 2.7$~Hz/K$^9$. $\alpha_{\mathrm{TLS}}$ is smaller than that for \Er{} nanoparticles suggesting a reduced concentration in structural defects, likely to be also due to \Nd{} nanoparticles larger size and annealing temperature. 

\subsection{Spectral diffusion in erbium  nanoparticles}\label{sec:3ppe}

\begin{figure}[ht]
\centering
\includegraphics[width=1.0\columnwidth]{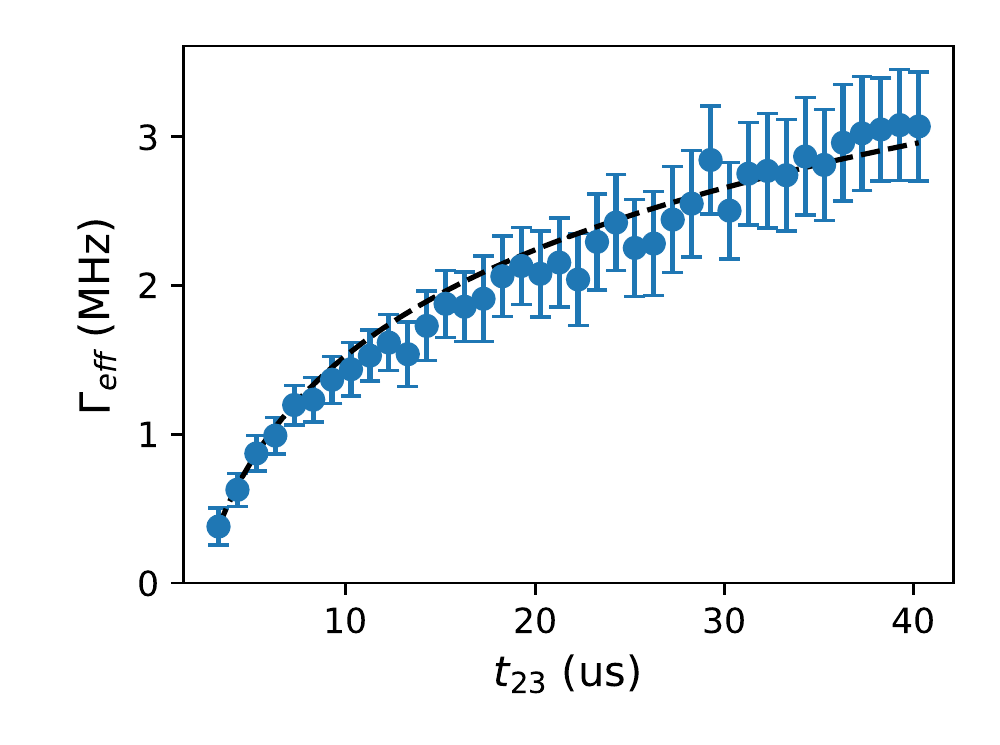}
\caption{Effective homogeneous linewidth of \ErYO{} nanoparticles as a function of $t_{23}$ in a three-pulse photon echo sequence. Dashed line is a fit to Eq.~\ref{eq:sd}, giving $\Gamma (t_0) = 379$~KHz and $\gamma = 2.3$~MHz. Error bars are one standard deviation of parameter estimate from fit.}
\label{fig:3ppe}
\end{figure}

We now investigate spectral diffusion in \ErYO{} nanoparticles via the three-pulse photon echo (3PPE) technique, which allows us to probe the broadening of the homogeneous linewidth on a time-scale that is much longer than the coherence time. To quantify the broadening $\Gamma_{\textrm{eff}}$ as a function of time, we keep the separation between the first two pulses in the 3PPE technique $t_{12}$ constant at 150~ns, while varying the separation between pulses two and three $t_{23}$ from $3$~$\mu$s to $40$~$\mu$s. The decay of the echo amplitude in such a 3PPE technique is given by \cite{Bottger2006a}:

\begin{equation}
E = E_0 \exp \left( \frac{-t_{23}}{T_1} \right) \exp (-2t_{12} \pi \Gamma_{\mathrm{eff}}),
\label{eq:3ppe}
\end{equation}
where $T_1$ is the excited state lifetime. We have $T_1 = 13.68~$ms (as measured above), and $3~\mu$s $\leq t_{23} \leq 40~\mu$s. Hence, the first factor is negligible for our parameters and all decay in echo amplitude is expected to come from a broadening of $\Gamma_{\mathrm{eff}}$. We begin by recording the decay in the echo amplitude as a function of $t_{23}$. For the first point of $t_{23} = 3~\mu$s, we set $\Gamma_{\mathrm{eff}}$ to the value of $\Gamma_h$ that was obtained from the 2PPE technique, that is $\Gamma_{\mathrm{eff}} = 379~$kHz. By estimating the relative decay in echo amplitude for all subsequent $t_{23}$, we extract $\Gamma_{\mathrm{eff}}$ as a function of $t_{23}$ that is shown in Fig. \ref{fig:3ppe}. For systems coupled to two-level systems (TLS), the $\Gamma_{\mathrm{eff}}$ is expected to have the following behaviour \cite{Veissier2016}:
\begin{equation}
\Gamma_{\mathrm{eff}}(t_{12},t_{23}) = \Gamma (t_0) + \gamma_{sd} \log_{10} \left( \frac{t_{23}}{t_0} \right),
\label{eq:sd}
\end{equation}
where $t_0$ is the minimum value of $t_{12} + t_{23}$ and $\gamma_{sd}$ is a coupling coefficient. For our case, $t_0 \approx 3~\mu$s and $\Gamma_0 = 379~$kHz. Fitting the $\Gamma_{\mathrm{eff}}$ that we obtained above to Eq.(\ref{eq:sd}), we obtain $\gamma_{sd} = 2.3$~MHz. This is a relatively high value for $\Gamma_{sd}$. For instance, Veissier et al. \cite{Veissier2016} measured $\Gamma_{sd}=0.4$ MHz for an erbium doped fiber at a temperature of 0.7 K. Further studies of spectral diffusion as a function of temperature are needed to investigate this.

\subsection{Homogeneous linewidth versus magnetic field in neodymium nanoparticles} \label{sec:bfield}

\begin{figure}[ht]
\centering
\includegraphics[width=1.00\columnwidth]{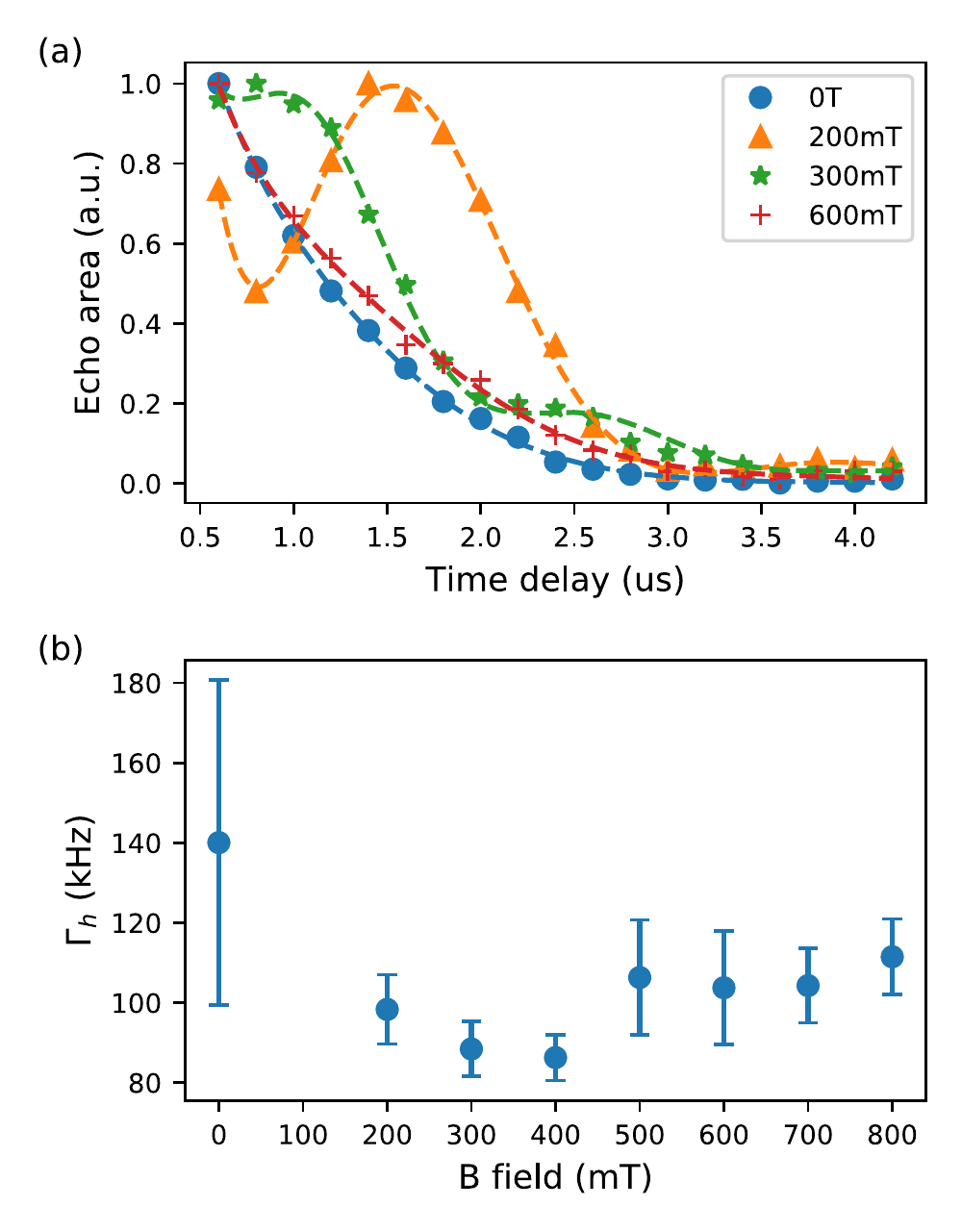}
\caption{(a) Photon echo decay at 0 T (blue), 200 mT (red), 300 mT (green) and 600 mT (magenta) measured in \Nd{} nanoparticles. Magnetic field dependence of the (b) optical homogeneous linewidth. Error bars represent 95\% confidence interval obtained from the fit.}
\label{fig:bfield}
\end{figure}

Kramers' ions with electron spin could be more sensitive to applied magnetic fields than other ions, so for quantum applications the T$_2$ dependence of the magnetic field is of relevance. Coherence measurements were taken at different magnetic fields for \Nd{}nanoparticles. In Fig.~ \ref{fig:bfield}(a) the echo area is plotted vs the magnetic field and a clear modulation can be seen on the echo decays when certain magnetic fields are applied. To extract the coherence time and the modulation frequency, the echo decays were fitted to the following equation \cite{Serrano2019}:

\begin{equation}
I(t) = I \times e^{(-4t/T_2)} \times[1+(m\times cos^2(\omega t/2))]^2
\end{equation}

\noindent where $T_2$
is the optical coherence time, $m$ is the modulation amplitude and $\omega$ is the modulation frequency.

The magnetic field dependence of the optical $\Gamma_h$ is shown in Fig.~\ref{fig:bfield}(b). The modulation frequency ranges between 200 kHz - 900 kHz. Similar modulation has been measured by Zhong \textit{et al} in Nd$^{3+}$:YVO$_4$, and it was attributed to superhyperfine interaction between the Nd$^{3+}$ electronic spin and the surrounding $Y^{3+}$ nuclear spin \cite{Zhong2018}.
As can be seen in the echo decay plots in Fig.~\ref{fig:bfield}(a), no modulation was observed at 0 T, which leads to the high error bars of the fit parameters.
While the superhyperfine modulation is strongly dependent on the magnitude of the magnetic field, as predicted by the Hamiltonian described in the supplemental material of \cite{Zhong2018}, no increase in the modulation frequency was observed after 400 mT due to the limited sampling.
The saturation of the homogeneous linewidth, $\Gamma_h$, around 100 kHz in Fig.~\ref{fig:bfield}(b) can be attributed to the superhyperfine limit, where the dephasing is defined by the interaction between \Nd{} and the surrounding Yttrium nuclei \cite{Sun2002Jul,Car2018,Car2020}. Another effect that could contribute to such saturation is the tunneling two-level systems \cite{Phillips1972,Fukumori2020}, which causes fluctuations in the local environment leading to some dephasing. The results show that the optical homogeneous linewidth is relatively unaffected by magnetic fields, and in particular, remains reasonably narrow at zero magnetic field. This is favorable since it allows quantum applications to potentially run without magnetic field, simplifying the setup and avoiding level splitting of other species like qubit ions.

\section{Applications}\label{sec:applications}

\subsection{Single Qubit State Readout}\label{sec:readout}

Scalability in quantum computers (QC) necessitate a fast and high fidelity detection of single qubit states. This is challenging for QC schemes based on RE ions due to its naturally forbidden electric dipole transitions. However, there have been several attempts recently to utilize the Purcell enhancement effect to detect single RE ions \cite{Chen2020Oct,Kindem2020Apr,Zhong2018,Casabone2021}. With high enough enhancement, one can either directly detect the qubit state, or use a different neighbouring ion as a readout qubit \cite{Kindem2020Apr,Chen2020Oct,Walther2015}.
In order to avoid any decay of the qubit excited state due to cavity enhancement, we are considering the second approach for the discussion below. In this approach, one can detect the state of a qubit ion that is close to the readout ion, by exciting one of the spin qubit states to the optically exited state. This will shift the levels of the readout ion (or not if the other state was instead populated), using the dipole blockade mechanism \cite{Walther2015}. This means that state detection can be made by separating the photon counts for ion emitting vs. not emitting (only background).
To ensure that the readout time is not a limiting factor in terms of QC speed, the readout should take place on a time scale comparable to the time scale of the gate operation. Typical duration of proposed gate operations in RE systems is in the order of a few microseconds \cite{Kinos2021Nov}. Furthermore, the homogeneous linewidth of the readout ion, which becomes broader for shorter lifetime, should be narrower than the frequency shift attained due to the dipole-dipole interaction with the neighbouring qubit ion to be detected. This is necessary to obtain high contrast in the photon counts when reading the two qubit states.

In order to see how the parameters measured in this study affects the capacity of our system to readout a QC, we consider our current \NdYO{} system with Pr$^{3+}$ (500 ppm) as qubit and Nd$^{3+}$ (100 ppm) as a candidate readout ion. We then discuss an ideal system with Eu$^{3+}$ as qubit.

The higher concentration of Pr$^{3+}$ increases the chance of having strongly interacting qubit ions in proximity to the readout ions, which is necessary for the dipole blockade mechanism to work. Given the data we measured, it would be useful to evaluate how good the Nd$^{3+}$ ion will perform in terms of readout fidelity.

When coupling an emitter to a cavity, the emitter's excited state lifetime is reduced to $T_c=T_1/(C+1)$, where $T_1$ is the natural lifetime, $T_c$ is the Purcell enhanced lifetime, and $C$ is the Purcell factor given by \cite{Purcell1946}
\begin{equation}\label{eq:purcell}
C = \zeta \frac{3\lambda^3}{4\pi^2}\frac{Q}{V_m}
\end{equation}
where $\lambda$ is the emission wavelength, $Q$ the quality factor of the resonator, $V_m$ its mode volume and $\zeta$ the branching ratio of the respective transition.

Nanoparticles that are investigated in this work are well suited to be integrated into fiber-based microcavities. However, scattering losses due to the particles are one of the key challenges to obtain high Q values. In particular, for the 380 nm Nd$^{3+}$ doped particles, the expected scattering loss is $\sim$ 8$\%$ per pass at 892 nm wavelength. Particles of 80 nm size have been realized \cite{Casabone2021}, and if we instead consider those, scattering losses will be reduced to $\sim$ 5 ppm. At this level, a loaded cavity finesse of $3\times10^5$ could be obtained using cavity mirrors with a total transmission of 10 ppm \cite{Rochau2021Jul}. For a loaded cavity with such a finesse, a Q $\sim$ $2 \times 10^6$ and $V_m$ $\sim$ 5$\lambda^3$ can be reached. From Eq. \ref{eq:purcell}, this corresponds to a Purcell enhancement C $\sim$ 5000 and an enhanced lifetime of $T_c \sim$ 100 ns.
Similarly, a Q $\sim$ $10^5$ and $V_m$ $\sim$ 5$\lambda^3$ can be reached for a cavity loaded with the 150 nm Er$^{3+}$ doped particles investigated here using similar cavity mirrors. Reducing the particle size to 100 nm would reduce the scattering losses and consequently increase Q by a factor of 3. Assuming an open cavity with those parameters, this would give C $\sim$ 5500 and $T_c \sim$ 2.5 $\mu$s for Er$^{3+}$. In order to reduce Er$^{3+}$ excited-state lifetime down to $\sim$ 100 ns, which is needed to obtain fast and reliable readout at a time scale comparable to the gate operation time as discussed earlier, a finesse of $4\times10^6$ and a cavity length of $\sim$ 1$\lambda$ will be required, which is quite challenging to obtain. Here, we will only consider Nd$^{3+}$ as a candidate readout ion with its 100 ns achievable enhanced lifetime.

Assuming an open cavity with the parameters discussed above, this would yield an enhanced lifetime of 100 ns for the envisioned readout ion Nd$^{3+}$. A detection rate of $10^6$ photons/s would be expected, given a reasonable assumption of 10 \% for the optical efficiency. Since the fiber is used both as a cavity mirror and as an outcoupler, a high collection efficiency would be expected, limited mainly by mode mismatching between the cavity and the fiber.

Debnath \textit{et al} proposed a protocol based on Bayesian analysis of the detected photons from a readout ion to extract information about the state of the qubit ion \cite{Debnath2020Jul}. In this protocol, the detection rate (1 MHz in our case) replaces the decay rate. To estimate the readout fidelity of our system based on this protocol, we assume an interaction shift (dipole blockade) of about 5 MHz, compared to the estimated 1.6 MHz homogeneous linewidth of the cavity enhanced lifetime of 100 ns. Considering a qubit excited-state emission rate of $\sim$ 140 times the detection rate ($T_1\sim$ 140 $\mu$s was measured for Pr$^{3+}$:$\YO{}$ \cite{Serrano2019}), the protocol estimates a readout fidelity of $\sim 94\%$ after about 10 $\mu$s. 
It should be noted that an interaction shift of 10 - 100 MHz is consistent with shift between ions separated by a few nanometers, which is a typical separation between RE ions at our doping concentrations. If we consider Eu$^{3+}$ as a candidate qubit instead, with an excited-state lifetime of $\sim$2 ms, the fidelity scales up to $\sim$ 99$\%$ after 10 $\mu$s \cite{Zhong2015Jan}). This fidelity is limited by the finite lifetime of the qubit ion, since if it decays spontaneously during the detection, it will cause a loss of the state information.  However, this problem can be mitigated by using one qubit as a dedicated buffer to the readout ions, to which the state of any other qubit that should be readout can be transferred as described in details in ref. \cite{Walther2015}. This way, the state information of the qubit will be protected from decay events by the buffer ion. By cycling a few times between the qubit and the buffer ion, the readout fidelity can be improved significantly to 99.9\%. This would be at the cost of extending the detection duration to $\sim$ 25-40 $\mu$s, only limited by the qubit decay during the transfer pulses for projecting the qubit state onto the buffer ion.

\subsection{Source of Fourier-limited single photons}\label{sec:single_photons}

Single erbium ions would be ideal candidates for use as quantum nodes in a quantum network due to their emission at telecommunication wavelengths. Furthermore, erbium ions that are spatially close to each other in the solid-state matrix allow for multi-qubit operations, which is highly desirable in a quantum network. A key requirement for using single-emitters as quantum nodes is that the emitted photons should be indistinguishable. In the spectral domain, this means that the photons should be Fourier-transform-limited. To extract Fourier-limited photons from a single erbium ion, we need to have $T_2=2T_1$, where $T_1$ is the lifetime of the excited state, and $T_2$ is the coherence time. For the values we have measured for \ErYO{} nanoparticles in this work at a temperature of 3~K, which is $T_1=13.7$~ms and $T_2=839$~ns, we have $T_2 \ll 2T_1$. For measurements in quantum networks where photons from different remote erbium emitters have to interfere at a beam splitter, it is not only necessary that the instantaneous linewidth is lifetime limited, but also that the central frequency does not drift such that the frequency overlap between the two photons remains high. This means that long term spectral diffusion effects have to be taken into account, such that the relevant parameter would be in that case $T_2^*$. Note however that slow drifts could be compensated for using e.g. external electric fields.

To satisfy the condition for indistinguishable photons, we would need to either increase $T_2$ (or $T_2^*$) or decrease $T_1$. As discussed above, the $T_2$ at low temperatures in the current samples is limited due to spectral diffusion arising from coupling to neighbouring two-level systems (TLS). TLS behaviour is characteristic of amorphous materials \cite{Veissier2016}, and could be arising in our system due to the presence of defects and the poor crystalline quality of our nanoparticles. As the contribution from TLS to homogeneous linewidth depends on temperature, coherence times can be significantly improved by cooling the samples further. In our current samples, the homogeneous linewidth reduces by 121~kHz per kelvin and reaches a value of 10~kHz ($T_2 \approx 32~\mu$s) when extrapolated to 0~K. The effect of TLS can also be minimized by annealing the samples at high temperatures ($>1200 ^\circ$C), but this also leads to an increase in the average size of our nanoparticles, which would make it difficult to integrate them into microcavities due to higher scattering losses. An alternative method that involves starting with bigger higher quality nanoparticles and then chemically etching them down to the required size has shown promising results \cite{Liu2020}. By chemically etching 400 nm \YO{} nanoparticles doped with europium down to 150 nm, they were able to demonstrate a $T_2$ of $10$ $\mu$s. Recent results obtained with \YO{} ceramics at mK temperatures have shown a sub kHz optical homogeneous linewidth \cite{Fukumori2020}.

Nonetheless, it is clear that both for the purpose of producing indistinguishable photons and also for making the emission efficient, we would need to reduce additionally the $T_1$. This can be achieved by coupling the emitter to an optical cavity and utilizing the Purcell effect as described above in Eq. \ref{eq:purcell} \cite{Casabone2021}. The \ErYO{} nanoparticles investigated in this work with an average size of 150~nm introduce a scattering loss of $\approx 35$~ppm. To efficiently extract photons emitted by ions in our nanoparticle coupled to a microcavity, this scattering loss needs to be smaller than all other losses in the cavity. For instance, our current nanoparticles can be efficiently coupled to cavities with $Q \sim 10^6$ and $V_m \sim 5~\lambda^3$. Using a lower bound $\zeta > 0.13$ as measured in this work, we expect a Purcell enhancement  $C > 2,000$, which corresponds to $T_c < 7~\mu$s. To produce indistinguishable photons, we would then need $T_2 \sim 14~\mu$s, which could be achieved both by improved sample fabrication as well as by cooling the samples further. Alternately, if the nanoparticles can be made with an average diameter of $\approx 100~$nm, then the scattering losses drop to $\approx 4~$ppm, in which case it would be possible to use cavities with $Q \sim 10^7$ that would result in $C \sim 20,000$ and $T_c \sim 700~$ns. In this case, $T_2 \sim 1.4~\mu$s would be sufficient to produce indistinguishable photons, which could be reached with the samples studied in this work by cooling them to $\sim 1.5~$K.

\section{Conclusion}
We have investigated the optical coherence properties of \Er{} and \Nd{} in \YO{} ceramics and nanoparticles at cryogenic temperatures. An inhomogeneous linewidth of 10.7 GHz and 8.2 GHz were measured for the \Ertransition{} transition in \Er{} and \Ndtransition{} transition in \Nd{}, respectively. Furthermore, the lifetime of the excited for the two transitions was measured to be 13.68 ms and 540 $\mu$s for the two aforementioned transitions, respectively. The two-pulse technique method was used to measure the homogeneous linewidth for the two ions at different temperatures. For \Er{} transition, a homogeneous linewidth of 379 kHz was measured at a base temperature of 3 K, while a linewidth of 62 kHz was measured in \Nd{} at a base temperature of 1.6 K. Time-dependent broadening of the homogeneous linewidth due to spectral diffusion was investigated in \Er{} using three-pulse photon echo technique, and a TLS coupling constant of 2.3 MHz was obtained. 

Finally, we propose two quantum application, single qubit-state readout and a Fourier-limited single photon source, for which the investigated ion are expected to perform well when combined with fiber-based microcavities, where a strong Purcell enhancement can be reached. In order to reach the necessary Purcell enhancement levels, cavities with quality factor in the range of 10$^6$-10$^7$ and a mode volume in the order of $\sim$ 5 $\lambda^3$ are required.

\section{\label{sec:level1} ACKNOWLEDGMENTS}
This research was supported by the European Union's Horizon 2020 research and innovation program no. 712721 (NanOQTech), no. 820391 (SQUARE), the Swedish Research Council (2021-03755), the Wallenberg Center for Quantum Technology (WACQT) funded by The Knut and Alice Wallenberg Foundation (KAW 2017.0449). ICFO acknowledges financial support from the government of Spain (PID 2019-106850RB-I00 (QRN) and Severo Ochoa CEX2019-000910-S, funded by MCIN/AEI/10.13039/501100011033), from MCIN with funding from European Union NextGenerationEU (PRTR-C17.I1), from Fundaci\'{o} ́Cellex, Fundaci\'{o} ́Mir-Puig, and from Generalitat de Catalunya (CERCA, AGAUR).

\bibliographystyle{apsrev4-1}
\bibliography{NdEr_v2}

\end{document}